\documentclass[11pt]{article}
\usepackage{amsmath}
\usepackage{amscd}
\usepackage{amssymb}
\usepackage{graphicx}
\usepackage{diagbox}
\usepackage{ifpdf}
\usepackage{fancyhdr}
\usepackage{hyperref}
\usepackage{multirow}
\usepackage{color}
\usepackage{listings}

\renewcommand{\baselinestretch}{1.3}
  \renewcommand{\arraystretch}{1.1}
\begin{document}

 \title{New Algorithms for Subset Sum Problem}

  \author{Zhengjun Cao$^{1}$,  \quad  Lihua Liu$^{2,*}$}
  \footnotetext{$^{1}$Department of Mathematics, Shanghai University, Shanghai, 200444,  China. \\
    $^2$Department of Mathematics, Shanghai Maritime University, Shanghai, 201306,  China. \,  $^*$\,\textsf{lhliu@shmtu.edu.cn} }

 \date{}\maketitle

\begin{quotation}
 \noindent\textbf{Abstract}. Given a set (or multiset) $S$ of $n$ numbers and a target number $t$, the subset sum
problem is to decide if there is a subset of $S$ that sums up to $t$.
There are several  methods for solving this problem, including exhaustive search, divide-and-conquer method, and Bellman's dynamic programming method.
However, none of them could generate universal and light code.
In this paper, we present a new deterministic algorithm based on a novel data arrangement, which could generate such code and return all solutions. If $n$ is small enough, it is efficient for usual purpose. We also present  a probabilistic version with one-sided error and a
 greedy algorithm which could generate a solution with minimized variance.


  \end{quotation}

\section{Introduction}

In computer science the subset sum problem is that: given a set (or multiset) of numbers, is there a non-empty subset whose sum is equal to a given number? The problem is generally expressed as follows:
    given numbers $w_1, \cdots, w_n$ along with a target number $t$,  the task is to determine whether there exists a subset $X\subset \{1, \dots, n\}$ such that  $w(X) :=\sum_{i\in X} w_i=t$.   Such an $X$ is referred to as a solution. For example, $t=5, w_1=1, w_2=2, w_3=3, w_4=4$. Clearly, $5=1+4=2+3$. It has two solutions.
 This problem is related to knapsack problem  and has many applications \cite{Dantzig57,Faaland73,Pisinger99,Kellerer04,Tran11,Uno12}.

 In 1956, Bellman \cite{Bellman56} introduced the dynamic programming method for subset sum, which was revisited \cite{Pferschy99,Pisinger03}.
  Let $A=\{w_1, w_2, \cdots, w_n\}$ and $A_j=\{w_1, w_2, \cdots, w_j\}$, $1\leq j\leq n$.
  Let $T[A_j, t]=1$ if $t$ can be expressed as the sum of some members of $A_j$, otherwise $T[A_j, t]=0$.
    We have the following recurrence formula
$$ T[A_j, t]= \text{max}\{T[A_{j-1}, t], T[A_{j-1}, t-w_j]\} \eqno(1) $$
 It is claimed that Bellman's method solves subset sum problem  in $O(nt)$ time. In fact,
 \begin{align*}
  T[A_j, t]= \text{max}&\{T[A_{j-1}, t], T[A_{j-1}, t-\textcolor[rgb]{1,0,0}{w_j}]\}\hspace*{6mm}(\text{1 addition})\\
  =\text{max}&\{T[A_{j-2}, t], T[A_{j-2}, t-\textcolor[rgb]{0,0,1}{w_{j-1}}],
   T[A_{j-2}, t-w_j], T[A_{j-2}, t-w_j-\textcolor[rgb]{0,0,1}{w_{j-1}}]\} \hspace*{2mm}(\text{2 additions})\\
  =\text{max}&\{T[A_{j-3}, t], T[A_{j-3}, t-\textcolor[rgb]{1,0,0}{w_{j-2}}], \\
  & T[A_{j-3}, t-w_{j-1}], T[A_{j-3}, t-w_{j-1}-\textcolor[rgb]{1,0,0}{w_{j-2}}] \\
  &T[A_{j-3}, t-w_j], T[A_{j-3}, t-w_j-\textcolor[rgb]{1,0,0}{w_{j-2}}],\\
  & T[A_{j-3}, t-w_j-w_{j-1}], T[A_{j-3}, t-w_j-w_{j-1}-\textcolor[rgb]{1,0,0}{w_{j-2}}]\} \hspace*{6mm}(\text{4 additions})\\
  =\text{max}&\{ \cdots \}.
 \end{align*}
 \textit{At the worst case}, the method needs $O(2^n)$ additions and stores $nt$ numbers of $1$ or $0$. Note that it is somewhat difficult to transform the recurrence method into  universal and light code because of its naive data arrangement.

  There are several methods to solve subset sum problem, such as \cite{Galil91,Kellerer03,Hamidoune08,Bringmann17,Koiliaris17}.
    Very recently,  Koiliaris and Xu \cite{Koiliaris17} have introduced a new version of divide-and-conquer method for subset sum problem.  It aims to  improve the ``conquer" step by taking advantage of the structure of the related sets.
 In 2017, Bringmann \cite{Bringmann17} presented a new algorithm for subset sum. For sets $A, B$ of non-negative integers, it defines
 $$A\oplus B =\{a+b\,|\, a\in A\cup \{0\}, b\in B\cup \{0\}\} \eqno(2) $$
 For any integer $t>0$, define
 $$A\oplus_tB:=(A\oplus B)\cap \{0, \cdots, t\} \eqno(3)$$
 For a given set $Z$,
 randomly partition it into
 $$Z = Z_1 \cup \cdots \cup Z_{k^2}$$
  i.e.,  assign any $z \in Z$
to a set $Z_i$ where $i$ is chosen independently and uniformly at random in $\{1, \cdots, k^2\}$. The basic subroutine of Bringmann's algorithm is the following.\vspace*{-3mm}

\noindent{\rule[-0.25\baselineskip]{\textwidth}{0.25mm}}

\noindent\textbf{ColorCoding}$(Z, t, k, \delta)$: Given a set $Z$ of positive integers, target $t$, size bound $k\geq 1$ and error probability $\delta>0$, we solve SUBSETSUM with solution size at most $k$.

1:  \textbf{for}  $j=1, \cdots, \lceil \log_{4/3}(1/\delta)\rceil$ \textbf{do}

2:   \hspace*{4mm} randomly partition $Z=Z_1\cup \cdots \cup Z_{k^2}$

3:   \hspace*{4mm} $S_j=Z_1\oplus_t \cdots \oplus_t Z_{k^2}$

4: \textbf{return} $\cup_jS_j$\vspace*{-1mm}

\noindent{\rule[0.55\baselineskip]{\textwidth}{0.25mm}}

Let $n_i=|Z_i|+1$ for $i=1, 2, \cdots, k^2$.
In the Step 3 of each round, this algorithm needs
$$n_1n_2+\text{min}\{n_1n_2, t\}n_3+\text{min}\{\text{min}\{n_1n_2, t\}n_3, t\}n_4\leq n_1n_2+t(n_3+\cdots+n_{k^2})\approx |Z|t.$$
Therefore, it needs $O(\lceil\log_{4/3}(1/\delta)\rceil |Z|t)$ additions totally. However, we should stress that the algorithm can not find all solutions.

We would like to point out  that none of these algorithms could generate concise and  amicable code. Moreover,
 these algorithms can not return all solutions. In this paper, we present a new algorithm based on a novel data arrangement, which could generate universal and light code and return all solutions. If $n$ is small enough, it is efficient for usual purpose. We also present a probabilistic version for returning a solution which runs in polynomial time with one-sided error, and a greedy algorithm which could generate a solution with minimized cardinal and variance.

 \section{A deterministic algorithm for subset sum}

 \subsection{Description of the new algorithm}

 The new algorithm  aims to find all solutions of the subset sum problem, $(w_1, \cdots, w_n; t)$.
 As we know the naive exhaust search method needs to compute $2^n-1$ values,
 \begin{align*}
 &w_1, w_2, \cdots, w_n,\\
 &w_1+w_2, w_1+w_3, \cdots, w_1+w_n, w_2+w_3, \cdots, w_{n-1}+w_n, \\
 &\cdots, \\
 &w_1+w_2+\cdots+w_n.
 \end{align*}
 These values are arranged naively. It could not generate light code.
 We find the following data arrangement is more heuristic.
$$t, t-\textcolor[rgb]{0,0,1}{w_1}, t-\textcolor[rgb]{1,0,0}{w_2}, t-w_1-\textcolor[rgb]{1,0,0}{w_2}, t-\textcolor[rgb]{0,0,1}{w_3}, t-w_1-\textcolor[rgb]{0,0,1}{w_3}, t-w_2-\textcolor[rgb]{0,0,1}{w_3}, t-w_1-w_2-\textcolor[rgb]{0,0,1}{w_3},$$
{\small$$t-\textcolor[rgb]{1,0,0}{w_4}, t-w_1-\textcolor[rgb]{1,0,0}{w_4}, t-w_2-\textcolor[rgb]{1,0,0}{w_4}, t-w_1-w_2-\textcolor[rgb]{1,0,0}{w_4}, t-w_3-\textcolor[rgb]{1,0,0}{w_4}, t-w_1-w_3-\textcolor[rgb]{1,0,0}{w_4}, t-w_2-w_3-\textcolor[rgb]{1,0,0}{w_4}, t-w_1-w_2-w_3-\textcolor[rgb]{1,0,0}{w_4}, \cdots$$}

Interestingly, each term in the above sequence can be written down using only its position $k$. For example,
\begin{itemize}
\item{} if $k=14$,  $2(k-1)=2(14-1)=26=(11010)_2$, and the term is $t-w_1-w_3-w_4$;
\item{} if $k=9$,  $2(k-1)=2(9-1)=16=(10000)_2$, and the term is $t-w_4$;
\item{} if $k=5$,  $2(k-1)=2(5-1)=8=(1000)_2$, and the term is $t-w_3$.
\end{itemize}

Based on this observation, the new algorithm can be described as follows.

\noindent{\rule[-0.25\baselineskip]{0.85\textwidth}{0.25mm}}

\textbf{Input}: $t, W=\{w_1, \cdots, w_n\}$.

\textbf{Output}: All solutions of $t$ with respect to $W$.

1:  Compute $t, t-w_1, t-w_2, t-w_1-w_2, \cdots, t-w_1-w_2-\cdots-w_n$.

2:  Find all positions of 0 in the above sequence.

3:  For each position $k$, compute the binary string $b_ib_{i-1}\cdots b_1b_0$ of $2(k-1)$,

\hspace*{5mm}and write down the solution $\{b_i\times w_i, b_{i-1}\times w_{i-1}, \cdots, b_1\times w_1\}$.

\noindent{\rule[0.55\baselineskip]{0.85\textwidth}{0.25mm}}

\subsection{The practical code}

We here present a light code for the algorithm, which is written in language ``Mathematica" because of its easily available and amicable windows.

{\small
\begin{verbatim}
BeginPackage["DiscreteMath`SubsetSum`"]
SubsetSum::usage = "Find all solutions for subset sum problem."
Begin["`Private`"]
SubsetSum[t_, W_List]:=
Module[{T,K,A,len,F},
      T={t}; 	
      For[i=1,i<=Length[W],i++,B=T-W[[i]];T=Join[T,B]];	
      K=Flatten[Position[T,0]];
      If[Length[K]>0,
         For[i=1,i<=Length[K],i++,		
             A=IntegerDigits[2(K[[i]]-1),2];
             len=Length[A];F={};
             For[j=1,j<=len,j++,If[A[[j]]==1,F=Append[F,W[[len-j]]]]];			
             Print[F]
            ],
        Print["Fail"]
        ]
      ]
End[]
EndPackage[]
\end{verbatim}}

Copy the above code and name the file as ``SubsetSum.m". Place the file into the folder

{\small
\verb|C:\Program Files\Wolfram Research\Mathematica\8.0\AddOns\LegacyPackages\DiscreteMath|
 }

In the interactive window for Mathematica, input ``$<<$ DiscreteMath`SubsetSum`" to upload the package. See the following Fig. 1 for the details.
\vspace*{3mm}

\hspace*{-8mm}\begin{minipage}{\linewidth}
\begin{center}
Figure 1:  Examples for the deterministic algorithm for  subset sum problem\vspace*{2mm}

\includegraphics[angle=0,height=7cm,width=9.5cm]{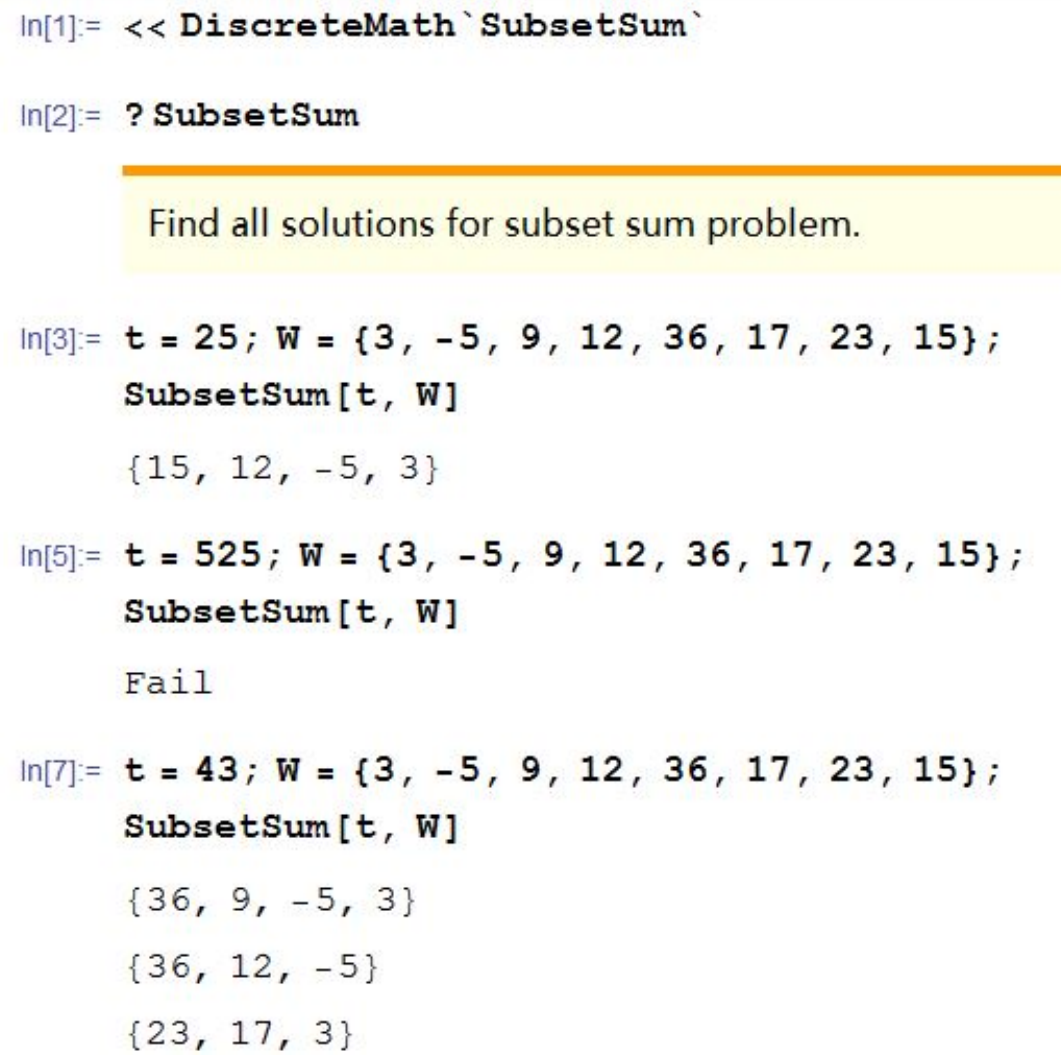}
\end{center}
 \end{minipage}

 \subsection{Analysis of the new algorithm}

The new algorithm is a variation of general exhaust search. Its novel data arrangement results directly  in the practical and light code.
It depends neither on any special property of the target number $t$, nor on the properties of the given sequence. So, it is universal for all cases.

 It needs $O(2^{n})$ basic arithmetic operations and stores $O(2^{n})$ numbers. Clearly, if $n$ is small enough, it works well. Otherwise, it becomes very hard because  subset sum  is a classical  NP-complete problem \cite{Karp72,Austrin16}. However, it is the first algorithm for subset sum problem which returns all solutions.

 \section{A probabilistic algorithm for subset sum}

If $n$ is large and only one solution is wanted,
one can randomly permutate the original sequence and truncate it into a short piece. Repeat the process many times.
The probabilistic algorithm can be described as follows.

\noindent{\rule[-0.25\baselineskip]{0.85\textwidth}{0.25mm}}

\textbf{Input}: $t, W=\{w_1, \cdots, w_n\}$.

\textbf{Output}: A solution of $t$ with respect to $W$, or the failure notation ``$\bot$".

1: Randomly permutate the sequence $w_1, \cdots, w_n$ and

 \hspace*{4mm} truncate it into a short piece $w_1', w_2', \cdots, w_k'$.

2:  Compute $t, t-w_1', t-w_2', t-w_1'-w_2', \cdots$.

3:  Find the first position of 0 in the above sequence. For the position $k$,

\hspace*{4mm} compute the binary string $b_ib_{i-1}\cdots b_1b_0$ of $2(k-1)$ and write down

\hspace*{4mm}  the solution $\{b_i\times w_i', b_{i-1}\times w_{i-1}', \cdots, b_1\times w_1'\}$.

4: If Step 3 fails, goto Step 1 and repeat the process for many times.

\noindent{\rule[0.55\baselineskip]{0.85\textwidth}{0.25mm}}

See the following figure 2 for some examples. \vspace*{3mm}

\hspace*{-5mm}\begin{minipage}{\linewidth}
\begin{center}
Fig. 2:  Examples for the probabilistic algorithm for subset sum problem  \vspace*{2mm}

\includegraphics[angle=0,height=9.5cm,width=15.5cm]{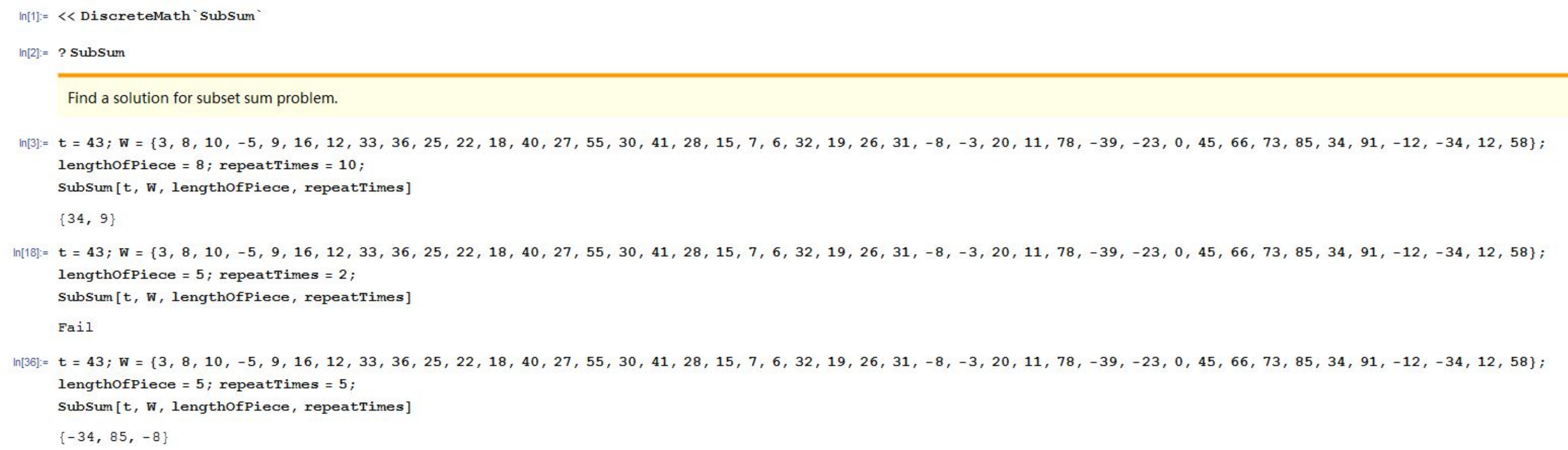}
\end{center}
 \end{minipage}\vspace*{2mm}

Here is the practical code.

{\small
\begin{verbatim}
BeginPackage["DiscreteMath`SubSum`"]
SubSum::usage = "Find a solution for subset sum problem."
Begin["`Private`"]
SubSum[t_, W_List, lengthOfPiece_, repeatTimes_]:=
Module[{n,a,V,B,T,k,A,len,F},
       n=Length[W];a=0;
       Do[V=Part[W,Union[Table[Random[Integer,{1, n}],{lengthOfPiece}]]]; 	
          T={t};
          For[j=1,j<=Length[V],j++, 		
              B=T-V[[j]]; T=Join[T,B];
              If[MemberQ[T,0],Break[]];
             ];
          If[MemberQ[T,0],
             k=First[Flatten[Position[T,0]]];
             A=IntegerDigits[2(k-1),2];
             len=Length[A];	F={};
             For[i=1,i<=Length[A],i++,If[A[[i]]==1,F=Append[F,V[[len-i]]]]];
             Print[F];a=1		
            ];
          If[a==1,Break[]],
          {repeatTimes}
          ];
        If[a==0,Print["Fail"]]
     ]
End[]
EndPackage[]
\end{verbatim}}

\section{A greedy algorithm for positive integers}

As we mentioned before, the deterministic algorithm needs $O(2^{n})$ basic arithmetic operations.
It has to store
$$t, t-w_1, t-w_2, t-w_1-w_2, t-w_3, t-w_1-w_3, t-w_2-w_3, t-w_1-w_2-w_3,
$$
$$ t-w_4, t-w_1-w_4, t-w_2-w_4, t-w_1-w_2-w_4, t-w_3-w_4, t-w_1-w_3-w_4,  \cdots. \eqno(4) $$
 If all elements of the given set $W$ are greater than 0, then there could be many negative integers in the above sequence. Clearly, these negative integers can be immediately deleted in each round if only one solution is wanted.
 \vspace*{-2mm}

\noindent{\rule[-0.25\baselineskip]{\textwidth}{0.25mm}}

\textbf{Input}: $t, W=\{w_1, \cdots, w_n\}, w_i>0, i=1, \cdots, n$, and $\ell$, a bound for rounds.

\textbf{Output}: A solution of $t$ with respect to $W$, which is of minimized variance, or the failure notation ``$\bot$".

1: Sort the sequence $w_1, \cdots, w_n$ decreasingly.

2:  Compute the sequence $t, t-w_1, t-w_2, t-w_1-w_2, \cdots $.  Delete all negative integers and \\
 \hspace*{8mm} merge those multiple elements, and sort the new sequence decreasingly in each round.

3:  Once 0 is found in the $k$-th round,  return $w_k$. $t\leftarrow t-w_k$.   \\
\hspace*{8mm} If $t\not=0$ and the round $k<\ell$, goto Step 2.

4:  If $t=0$, return the special solution which is of minimized variance. \\
 \hspace*{8mm}  Otherwise, return ``$\bot$".

\noindent{\rule[0.55\baselineskip]{\textwidth}{0.25mm}}

In order to further reduce the computational cost, we suggest
arrange the integers in $W$ decreasingly, i.e., $w_1\geq w_2\geq \cdots\geq w_n$. In each round, those multiple elements can be merged immediately. Thus, the final solution is of \textit{minimized cardinal and variance}. Of course, it needs least arithmetic operations.
See the figure 3 for an illustration of the greedy algorithm. \vspace*{4mm}

 \hspace*{-5mm}\begin{minipage}{\linewidth}
\begin{center}
Fig. 3:  A greedy algorithm for positive integers   \vspace*{2mm}

\includegraphics[angle=0,height=10cm,width=13.5cm]{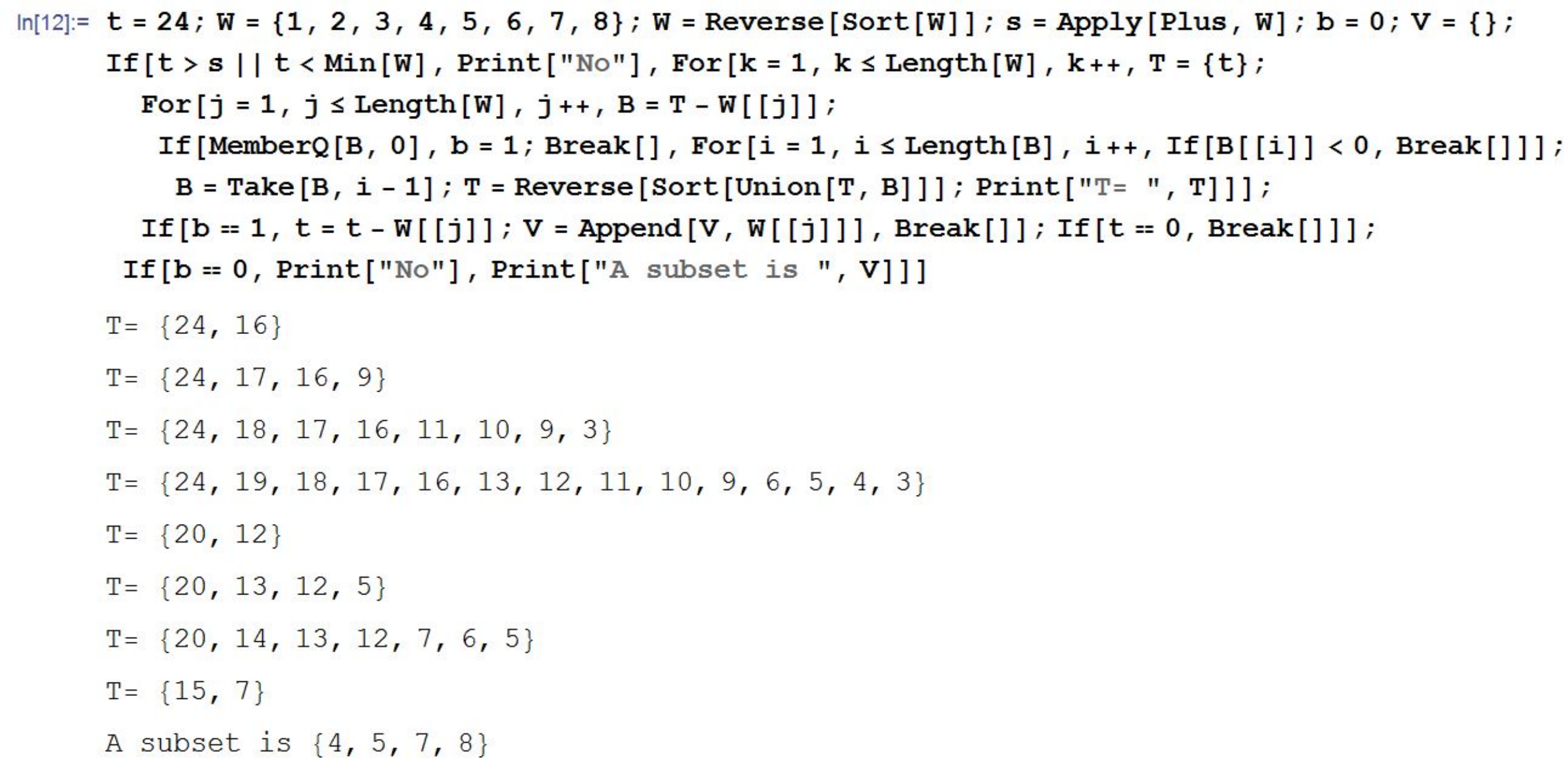}
\end{center}
 \end{minipage}\vspace*{2mm}

\begin{center}{\small
 \begin{tabular}{|l|c|}
   \hline
  $t=24, W=\{1,2,3,4,5,6,7,8\}$  & Sample variance $s^2$ \\ \hline
   $\{8, 7, 6, 2, 1\}$  & 9.7000 \\
    $\{8, 7, 5, 3, 1\}$  & 8.2000 \\
  $\{8, 6, 4, 3, 2, 1\}$ & 6.8000 \\
   $\{8, 6, 5, 4, 1\}$ & 6.7000 \\
  $\{8, 7, 4, 3, 2\}$ & 6.7000 \\
  $\{8, 6, 5, 3, 2\}$ & 5.7000 \\
 $\{{7,6,5,3,2,1}\}$ & 5.6000 \\
    $\{8, 7, 6, 3\}$  & 4.6667 \\
   $\{{7, 6, 5, 4, 2}\}$  & 3.7000 \\
      \textcolor[rgb]{1,0,0}{$\{8, 7, 5, 4\}$}  & \textcolor[rgb]{1,0,0}{3.3333}\\
   \hline
 \end{tabular}}\end{center}

The complexity of the greedy algorithm depends either on the position of least integer of the solution with minimized variance, or on the number of negative integers in the basic sequence.
The following figure 4 depicts some experimental results. The randomly picked integers are of 20-bit length.
It shows that $n=28$ is a practical threshold value for the deterministic algorithm  on PC, which requires $O(2^{28})$ arithmetic operations. However,
for a \textit{random set} $W$, the threshold value for the greedy algorithm is expected to be greater than $64$.

\vspace*{3mm}

\hspace*{-5mm}\begin{minipage}{\linewidth}
\begin{center}
Fig. 4:  Experimental results   \vspace*{2mm}

\includegraphics[angle=0,height=6.5cm,width=13.5cm]{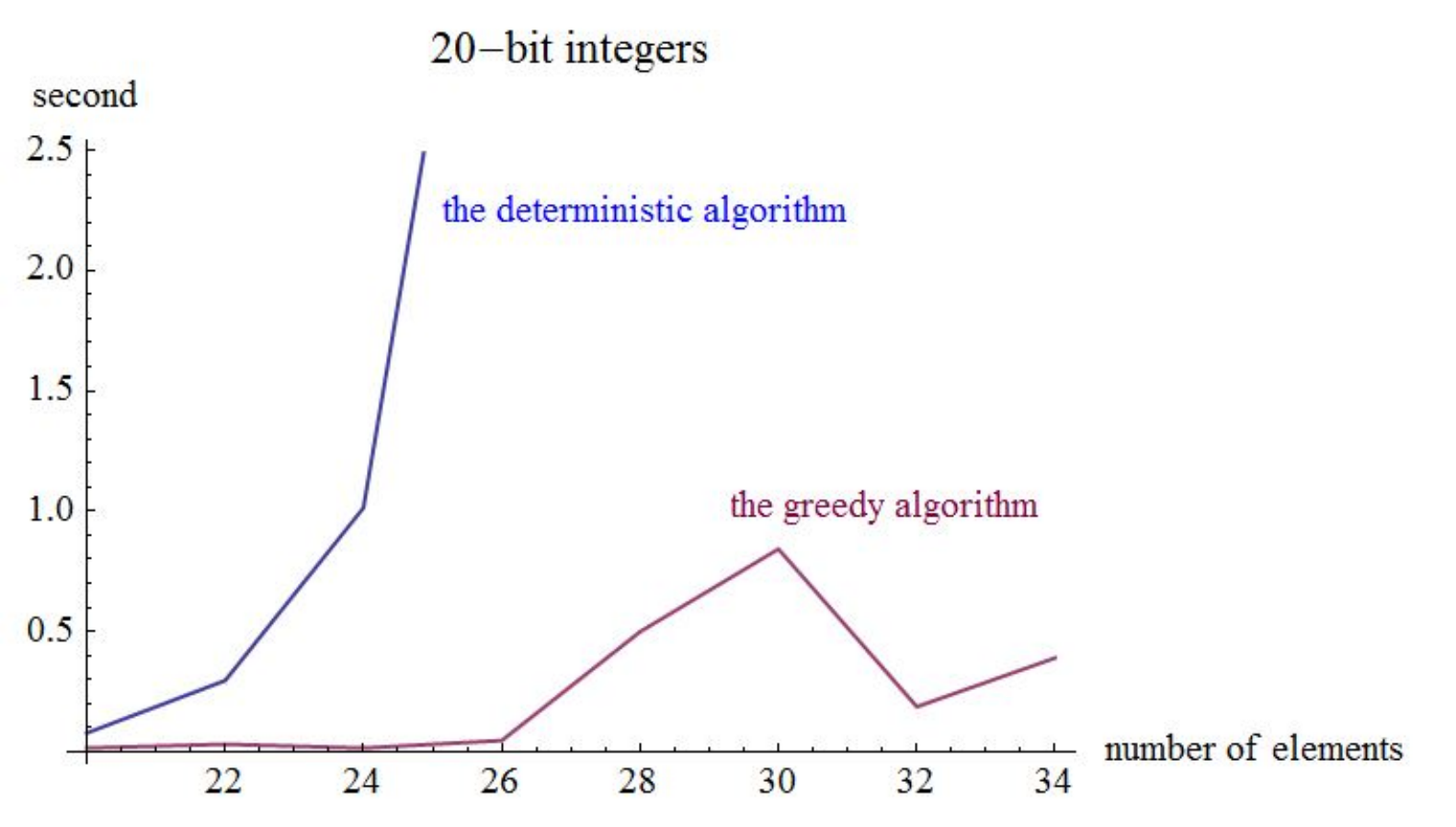}
\end{center}
 \end{minipage}

\section{Conclusion}

In this paper, we present three algorithms for subset sum problem which are based a novel data arrangement.
The deterministic is universal and practical which can find all solutions if the given set size is small enough.
We also propose a probabilistic version and a greedy algorithm for seeking a solution of subset sum problem.

\end{document}